\documentclass[doublespacing]{bmcart}

\usepackage[utf8]{inputenc} %unicode support

\usepackage{bm}% bold math
\usepackage{tikz}
\usepackage{url}
\usepackage{textcomp}

\usepackage{mathtools}
\usetikzlibrary{patterns,calc}

\usepackage{color}

\startlocaldefs
\endlocaldefs

\begin{document}

%%% Start of article front matter
\begin{frontmatter}

\begin{fmbox}
\dochead{Research}


\title{A Systematic Framework of Modelling Epidemics on Temporal Networks}

%%%%%%%%%%%%%%%%%%%%%%%%%%%%%%%%%%%%%%%%%%%%%%
%%                                          %%
%% Enter the authors here                   %%
%%                                          %%
%% Specify information, if available,       %%
%% in the form:                             %%
%%   <key>={<id1>,<id2>}                    %%
%%   <key>=                                 %%
%% Comment or delete the keys which are     %%
%% not used. Repeat \author command as much %%
%% as required.                             %%
%%                                          %%
%%%%%%%%%%%%%%%%%%%%%%%%%%%%%%%%%%%%%%%%%%%%%%

\author[
   addressref={aff1},                   % id's of addresses, e.g. {aff1,aff2}
   %corref={aff1},                       % id of corresponding address, if any
   %noteref={n1},                        % id's of article notes, if any
   email={r.humphries@umail.ucc.ie}   % email address
]{\inits{RH}\fnm{Rory} \snm{Humphries}}
\author[
   addressref={aff1},                   % id's of addresses, e.g. {aff1,aff2}
   %corref={aff1},                       % id of corresponding address, if any
   %noteref={n1},                        % id's of article notes, if any
   email={k.mulchrone@ucc.ie}   % email address
]{\inits{KM}\fnm{Kieran} \snm{Mulchrone}}
\author[
   addressref={aff2},
   %email={john.RS.Smith@cambridge.co.uk}
]{\inits{JT}\fnm{Jamie} \snm{Tratalos}}
\author[
   addressref={aff2},
   %email={john.RS.Smith@cambridge.co.uk}
]{\inits{SM}\fnm{Simon} \snm{More}}
\author[
   addressref={aff1},                   % id's of addresses, e.g. {aff1,aff2}
   %corref={aff1},                       % id of corresponding address, if any
   %noteref={n1},                        % id's of article notes, if any
   email={philipp.hoevel@ucc.ie}   % email address
]{\inits{PH}\fnm{Philipp} \snm{H\"ovel}}

%%%%%%%%%%%%%%%%%%%%%%%%%%%%%%%%%%%%%%%%%%%%%%
%%                                          %%
%% Enter the authors' addresses here        %%
%%                                          %%
%% Repeat \address commands as much as      %%
%% required.                                %%
%%                                          %%
%%%%%%%%%%%%%%%%%%%%%%%%%%%%%%%%%%%%%%%%%%%%%%

\address[id=aff1]{%                           % unique id
  \orgname{School of Mathematical Sciences, University College Cork}, % university, etc
  \street{Western Road},                     %
  \postcode{T12 XF64}                                % post or zip code
  \city{Cork},                             % city
  \cny{Ireland}                                    % country
}
\address[id=aff2]{%
\orgname{UCD Centre for Veterinary Epidemiology and Risk Analysis, UCD School of Veterinary Medicine, University College Dublin},
  \street{Belfield},
  \postcode{D04 W6F6}
  \city{Dublin},
  \cny{Ireland}
}

%%%%%%%%%%%%%%%%%%%%%%%%%%%%%%%%%%%%%%%%%%%%%%
%%                                          %%
%% Enter short notes here                   %%
%%                                          %%
%% Short notes will be after addresses      %%
%% on first page.                           %%
%%                                          %%
%%%%%%%%%%%%%%%%%%%%%%%%%%%%%%%%%%%%%%%%%%%%%%

%\note{Sample of title note}     % note to the article
%\note[id=n1]{Equal contributor} % note, connected to author

\end{fmbox}% comment this for two column layout

%%%%%%%%%%%%%%%%%%%%%%%%%%%%%%%%%%%%%%%%%%%%%%
%%                                          %%
%% The Abstract begins here                 %%
%%                                          %%
%% Please refer to the Instructions for     %%
%% authors on http://www.biomedcentral.com  %%
%% and include the section headings         %%
%% accordingly for your article type.       %%
%%                                          %%
%%%%%%%%%%%%%%%%%%%%%%%%%%%%%%%%%%%%%%%%%%%%%%

\begin{abstractbox}

\begin{abstract} % abstract
We present a modelling framework for the spreading of epidemics on temporal networks from which both the individual-based and pair-based models can be recovered. The proposed temporal pair-based model that is systematically derived from this framework offers an improvement over existing pair-based models by moving away from edge-centric descriptions while keeping the description concise and relatively simple. For the contagion process, we consider the Susceptible-Infected-Recovered  (SIR) model, which is realized on a network with time-varying edges. We show that the shift in perspective from individual-based to pair-based quantities enables exact modelling of Markovian epidemic processes on temporal tree networks. On arbitrary networks, the proposed pair-based model provides a substantial increase in accuracy at a low computational and conceptual cost compared to the individual-based model. From the pair-based model, we analytically find the condition necessary for an epidemic to occur, otherwise known as the epidemic threshold. Due to the fact that the SIR model has only one stable fixed point, which is the global non-infected state, we identify an epidemic by looking at the initial stability of the model.
\end{abstract}

%%%%%%%%%%%%%%%%%%%%%%%%%%%%%%%%%%%%%%%%%%%%%%
%%                                          %%
%% The keywords begin here                  %%
%%                                          %%
%% Put each keyword in separate \kwd{}.     %%
%%                                          %%
%%%%%%%%%%%%%%%%%%%%%%%%%%%%%%%%%%%%%%%%%%%%%%

%\begin{keyword}
%\kwd{sample}
%\kwd{article}
%\kwd{author}
%\end{keyword}

% MSC classifications codes, if any
%\begin{keyword}[class=AMS]
%\kwd[Primary ]{}
%\kwd{}
%\kwd[; secondary ]{}
%\end{keyword}

\end{abstractbox}
%
%\end{fmbox}% uncomment this for twcolumn layout

\end{frontmatter}

%%%%%%%%%%%%%%%%%%%%%%%%%%%%%%%%%%%%%%%%%%%%%%
%%                                          %%
%% The Main Body begins here                %%
%%                                          %%
%% Please refer to the instructions for     %%
%% authors on:                              %%
%% http://www.biomedcentral.com/info/authors%%
%% and include the section headings         %%
%% accordingly for your article type.       %%
%%                                          %%
%% See the Results and Discussion section   %%
%% for details on how to create sub-sections%%
%%                                          %%
%% use \cite{...} to cite references        %%
%%  \cite{koon} and                         %%
%%  \cite{oreg,khar,zvai,xjon,schn,pond}    %%
%%  \nocite{smith,marg,hunn,advi,koha,mouse}%%
%%                                          %%
%%%%%%%%%%%%%%%%%%%%%%%%%%%%%%%%%%%%%%%%%%%%%%

%%%%%%%%%%%%%%%%%%%%%%%%% start of article main body
% <put your article body there>

%%%%%%%%%%%%%%%%
%% Background %%
%%
%\section*{Content}
%Text and results for this section, as per the individual journal's instructions for authors. %\cite{koon,oreg,khar,zvai,xjon,schn,pond,smith,marg,hunn,advi,koha,mouse}

\section{Introduction}\label{sec:intro}
In recent years epidemiological modelling, along with many other fields, has seen renewed activity thanks to the emergence of network science \cite{Newman2018, Barabasi2016,ZHA20,MAS17}. Approaching these models from the view of complex coupled systems has shed new light onto spreading processes where the early black-box ordinary differential equation (ODE) models from Kermack and McKendrick had its limitations \cite{Keeling2005, Humphries2020}. These ODE models assume homogeneous mixing of the entire population, which may be an appropriate approximation for small communities. However, when attempting to model the spread of disease at a national or international level, they fail to capture how heterogeneities in both travel patterns and population distributions contribute to and affect the spread of disease. Epidemiological models on complex networks aim to solve this problem by moving away from averaged dynamics of populations and mean-field descriptions. Instead, the focus is on interactions between individuals or meta populations, where the spreading process is driven by contacts in the network \cite{yang_wang_epidemic_2003, sharkey_exact_2015, colizza2007reaction}.

There have been many improvements made in regards to network models, e.g., generalised multi-layer network structures or more specifically temporal networks that allow for the network structure to change with time \cite{MAS17,IAN17,KOH16,LEN16}. Temporal networks are a natural way of representing contacts and lead to an insightful interplay between the disease dynamics and the evolving network topology \cite{karrer_message_2010,  shrestha_message-passing_2015, lentz2013unfolding}. With the ever growing availability of mobility and contact data it has become easier to provide accurate and high-resolution data to inform network models. The results can be extremely useful tools for public-health bodies and other stakeholders \cite{GET19, TRATALOS2020105095, genois_can_2018}.

In previous works, a widely used epidemiological concept is the individual-based model \cite{Newman2018, valdano_analytical_2015, sharkey_deterministic_2011}. It assumes statistical independence in the state of each vertex.  A major problem associated with such a model is that it suffers quite badly from an echo chamber effect due to the fact that there is no memory of past interactions due to statistical independence. There have been efforts to ameliorate this problem by introducing memory at the level of each vertex's direct neighbours. These models referred to as contact-based \cite{koher_contact-based_2019} or pair-based \cite{frasca_discrete-time_2016} and have been shown to significantly reduce the echo chamber effect, depending on the underlying network structure. These two models differ in their initial approach. The contact-based model takes an edge-based perspective, which extends the message passing approach \cite{karrer_message_2010, kirkwood_statistical_1935}, and all dynamic equations are formulated in terms of edges. By contrast, the pair-based model keeps the vertex-based approach of the individual-based model and dynamic equations are in terms of vertices.

In this paper, we extend the Pair-Based (PB) model to a temporal setting giving a Temporal Pair-Based (TPB) Model. We show how it can be drastically reduced and simplified under a certain dynamical assumption \cite{sharkey_complete_2015}. We deal specifically with susceptible-infected-recovered (SIR) models. Once the TPB model is written in concise form, it is then possible to show that the contact-based model is equivalent to a linearised version of the TPB model. We then establish the conditions for an epidemic to occur according to the TPB model, also known as the epidemic threshold. We investigate how the TPB model performs on a number of synthetic and empirical networks and investigate what kind of network topologies work best with the TPB model.

The remainder of this paper is structured as follows: In Section~\ref{sec:sir_net_model}, we summarize the theoretical framework. Then, we address the calculation of an epidemic threshold in Section~\ref{sec:epi_thresh}, before presenting the main results in Section~\ref{sec:results}. Finally, we offer some conclusions in Section~\ref{sec:conclusions}.

\section{SIR Network Model}\label{sec:sir_net_model}
Let us consider a temporal network $\mathcal{G}=\{G_1,\dots,G_T\}$ to be a series of networks $G_t = (V, E_t)$, which all share the same vertex set $V$ but differ in their edge set $E_t$. The adjacency matrix for the network at time $t$ will be denoted by $A^{[t]}$, and $A^{[t]}_{ij}=1$ implies an edge between vertices $i$ and $j$ at time $t$.

\subsection{Reduced Master Equations}
\label{sec:rme}
Let $\Omega$ be the set of compartments in the model, that is, in the SIR model: $\Omega = \{S,I,R\}$. Let  $\mathbf{X}^n=\left( X_1^n, X_2^n,\dots,X_N^n \right)^T$ be the vector whose $i$-th element refers to the state of the $i$-th vertex in the network at time step $n$, thus $\mathbf{X}^n\in \Omega^{N}$ where $|V|=N$. The evolution of the disease is then described by the master equations \cite{gardiner_stochastic_2009},
\begin{equation}\label{eq:me}
P(\mathbf{X}^{n+1}) = \sum_{\mathbf{X}^{n}\in \Omega^{N}}P(\mathbf{X}^{n+1}|\mathbf{X}^{n})P(\mathbf{X}^{n}).
\end{equation}
In other words, we assume that the infection process is Markovian. $P(\mathbf{X}^{n+1})$ is the probability of the network being in the particular configuration of states given by $\mathbf{X}^{n+1}$ and $P(\mathbf{X}^{n+1} | \mathbf{X}^{n})$ is probability of the network moving from the configuration of states $\mathbf{X}^{n}$ to $\mathbf{X}^{n+1}$ between their respective time steps. These equations describe the entire process on the network. However, in order to progress the system forward one step in time, the probabilities of all combinations of state vectors must be found. This usually is not feasible for network processes with potentially billions of vertices as for the SIR process the total combination of states is given by $3^N$.

Instead, it is possible to describe the evolution of the disease using a system of Reduced Master Equations (RME) \cite{sharkey_deterministic_2011}, which describes the evolution of subsystems within the network, such as individual vertices, removing the need to obtain every possible combination of states. An important note is that these RMEs are in fact not themselves true master equations as they are not necessarily linear due to the fact that the transition rates of the subsystems are non-linear combinations of the transitions rates of the original system. However, we shall continue to use the term RME introduced by the author of \cite{sharkey_deterministic_2011}.

For notational convenience we use the following notation for the joint marginal probabilities,
\begin{align}
\begin{split}
P(X^{n}_{i_1},X^{n}_{i_2},\cdots ,X^{n}_{i_m})=& \langle X^{n}_{i_1}X^{n}_{i_2}\cdots X^{n}_{i_m}\rangle.
\end{split}
\end{align}
When we wish to specify a particular realisation of $X_i^n$, we denote it by $S_i^n$, $I_i^n$ or $R_i^n$ to imply $X_i^n=S$, $X_i^n=I$ or $X_i^n=R$ respectively. Employing this new notation we start with the RME which describes the evolution of individual vertices,
\begin{equation}\label{eq:rme}
\langle X^{n+1}_{i} \rangle = \sum_{X^n_i\in \Omega}\langle X^{n+1}_i|X_i^n \rangle \langle X_i^n \rangle.
\end{equation}
For SIR dynamics, the evolution of each vertex in each compartment is given as the following,
\begin{subequations}\label{eq:SIR:rme:ind}
\begin{align}
\langle S^{n+1}_i\rangle& = \langle S^{n}_i\rangle
-\langle I_i^{n+1}|S^{n}_i \rangle \langle S^{n}_i\rangle\\
\langle I^{n+1}_i\rangle& = \langle I^{n}_i\rangle
-\langle R_i^{n+1}|I^{n}_i \rangle\langle I^{n}_i\rangle + 
\langle I_i^{n+1}|S^{n}_i \rangle\langle S_i^n\rangle\\
\langle R^{n+1}_i\rangle&=\langle R^{n}_i\rangle + 
\langle R_i^{n+1}|I^{n}_i \rangle \langle I_i^n\rangle ,
\end{align}
\end{subequations}
where $\langle I_i^{n+1}|S^{n}_i \rangle$ reads the probability vertex $i$ is infected at time $n+1$ given it was susceptible at time $n$ and similarly for $\langle R_i^{n+1}|I^{n}_i \rangle$. Note that $\langle R_i^n\rangle$ can be recovered using the conservation of the probabilities $ \langle S_i^n\rangle + \langle I_i^n\rangle + \langle R_i^n\rangle = 1$. In order to compute the transition rates we define the following quantities, the probability of infection on contact $\beta$, the rate of recovery $\mu$. $A^{[n]}$ is the temporal adjacency matrix of the network on which the process is occurring. Following directly from \cite{frasca_discrete-time_2016}, the transition rates of moving from $S$ to $I$, and $I$ to $R$ are given by
\small{
\begin{subequations}\label{eq:SIR:tr:ind}
\begin{align}
\langle I_i^{n+1}|S^{n}_i \rangle =& 
\frac{1}{\langle S_i^n\rangle}\left[\beta\sum_{j_1\in V}A_{ij_1}^{[n]} \langle S_i^n I_{j_1}^n\rangle\right.
- \beta^2\sum_{j_1, j_2\in V} A_{ij_1}^{[n]}A_{ij_2}^{[n]}\langle S_i^n I_{j_1}^n I_{j_2}^n\rangle\nonumber\\
&+\dots \left.- (-\beta)^{N-1}\smashoperator{\sum_{j_1,\dots,j_{N-1}\in V}} A_{ij_1}^{[n]}\dots A_{ij_{N -1}}^{[n]}\langle S_i^n I_{j_1}^n \dots I_{j_{N-1}}^n\rangle\right]\label{eq:SIR:S:I}\\
\langle R_i^{n+1}|I^{n}_i \rangle =& \mu.\label{eq:SIR:I:R}
\end{align}
\end{subequations}
}

For the expression within the square brackets of Eq.~\eqref{eq:SIR:S:I}, the first term is the probability that vertex $i$ is infected by some other vertex in the network, however, double counts events. Each subsequent term accounts for double-counting and over-correcting in the previous. These equations describe the probabilistic SIR process on temporal networks. However, the system of equations is not closed as it lacks a description for their joint probabilities. There are a number of ways which this problem can be tackled, usually by making a number of numerical or dynamical approximations \cite{koher_contact-based_2019,yang_wang_epidemic_2003,valdano_analytical_2015,gomez_discrete-time_2010}. In the next sections we attempt to improve on and unify many existing approaches showing how they are derived from the system of RME's given by Eqs. \eqref{eq:SIR:rme:ind} and \eqref{eq:SIR:tr:ind}.

\subsection{Temporal Individual-based Model}\label{sec:TIB_model}
One of the most commonly used epidemiological models on networks is individual-based (IB) model, which is extended to the temporal setting in \cite{valdano_analytical_2015}. We refer to this extension as the temporal individual-based (TIB) model. This refers to the assumption of statistical independence of vertices or the mean field approximation, i.e., the factorisation $\langle X_{i_1}^nX_{i_2}^n\dots X_{i_M}^n\rangle=\langle X_{i_1}^n\rangle \langle X_{i_2}^n\rangle\dots\langle X_{i_M}^n\rangle$. By assuming this independence of vertices, Eq.~(\ref{eq:SIR:S:I}) simplifies to 
\small{
	\begin{subequations}
		\begin{align}
		\langle I_i^{n+1}|S^{n}_i \rangle =& \beta\sum_{j_1\in V}A_{ij_1}^{[n]} \langle I_{j_1}^n\rangle\nonumber\\
		&- \beta^2\sum_{j_1, j_2\in V} A_{ij_1}^{[n]}A_{ij_2}^{[n]}\langle I_{j_1}^n\rangle\langle I_{j_2}^n\rangle +\dots \nonumber\\
		&-(-\beta)^{N-1}\smashoperator{\sum_{j_1,\dots,j_{N-1}\in V}} A_{ij_1}^{[n]}\dots A_{ij_{N -1}}^{[n]}\langle I_{j_1}^n\rangle \dots \langle I_{j_{N-1}}^n\rangle,
		\end{align}
	\end{subequations}
}
which upon factorising can be concisely written as 
\begin{equation}\label{eq:SIR:TIB:S:I}
\langle I_i^{n+1}|S^{n}_i \rangle =
1 - \prod_{k\in V}\left(1-\beta A_{ki}^{[n]}\langle I_k^n\rangle\right).
\end{equation}
Upon substituting the transition rates $\langle I_i^{n+1}|S^{n}_i \rangle$ and $\langle R_i^{n+1}|I^{n}_i \rangle$ under the assumption of statistical independence the full TIB model is written as
{\small
	\begin{subequations}\label{eq:SIR:TIB:rme}
	\begin{align}
	\langle S^{n+1}_i\rangle& = \langle S^{n}_i\rangle
	\prod_{k\in V}\left(1-\beta A_{ki}^{[n]}\langle I_k^n\rangle\right)\\
	\langle I^{n+1}_i\rangle& = \langle I^{n}_i\rangle(1-\mu)+\langle S^{n}_i\rangle\left(1-
	\prod_{k\in V}\left(1-\beta A_{ki}^{[n]}\langle I_k^n\rangle\right)\right).
	\end{align}
	\end{subequations}}

	A
This model closes the equation \eqref{eq:SIR:S:I} at the level of vertices, thus ignoring all correlations with other vertices at previous times. However, ignoring all past correlations causes the model to suffer quite badly from an echo chamber effect \cite{shrestha_message-passing_2015}. This echo chamber has the effect of vertices artificially amplifying each other's probability of being infected $\langle I_i^n \rangle$ at each new time step, as the marginal probability of each vertex is highly correlated with the rest of the network and the factorisation of Eq.~\eqref{eq:SIR:S:I} means each vertex forgets its past interactions. As demonstrated in \cite{shrestha_message-passing_2015}, it is possible to show that in the absence of a recovered compartment, a static network of two linked vertices for non-zero initial conditions has probabilities of being infected which converge according to $\lim_{n\rightarrow\infty}\langle I_0^n\rangle=\lim_{n\rightarrow\infty}\langle I_1^n\rangle=1$ for the TIB model.

\subsection{Temporal Pair-based Model}\label{sec:TPB_model}

In contrast to the TIB-model, instead of assuming independence of vertices we can approximate the marginal probabilities in terms of combinations of lower order marginals using some form of moment closure \cite{frasca_discrete-time_2016}\cite{sharkey_complete_2015}. Here we make an equivalent assumption to that of the message passing approaches \cite{shrestha_message-passing_2015}\cite{karrer_message_2010}. We assume the network contains no time-respecting non-backtracking cycles, in other words, starting at some initial vertex $i$ that leaves via vertex $j$, there is no way to find a time-respecting path returning to this vertex that does not return via $j$. This is equivalent to a tree network when the temporal network is viewed in its static embedding of the supra-adjacency representation \cite{bianconi_multilayer_2018}. This allows us to write all higher order moments in Eq.~\eqref{eq:SIR:S:I} as a combination of pairs $\langle S_i^n I_k^n\rangle$. To show why this is possible, consider the three vertices $i,j,k$ connected by two edges through $i$. If conditional independence of these vertices is assumed given we have the state of $i$, then one can make the following assumption,
A
\begin{equation}\label{eq:mom_closure}
\langle X_i^nX_j^nX_k^n\rangle = \langle X_j^nX_k^n|X_i^n\rangle\langle X_i^n\rangle
= \frac{\langle X_i^nX_j^n\rangle\langle X_i^nX_k^n\rangle}{\langle X_i^n\rangle}.
\end{equation} 

This has the effect of assuming the network is tree-like in structure as it implies any interaction between vertices $j$ and $k$ must occur through vertex $i$, thus the process is exact on networks which contain no time-respecting non-backtracking cycles and otherwise provides an improved approximation of varying degree which depends on the true network structure. The result obtained in Eq.~\eqref{eq:mom_closure} is often referred to as the Kirkwood closure \cite{kirkwood_statistical_1935}. Under the assumption that the network is tree like, the following simplification is obtained for Eq.~\eqref{eq:SIR:S:I},
\begin{equation}\label{eq:SIR:TPB:S:I}
\langle I_i^{n+1}|S^{n}_i \rangle=1 - \prod_{k\in V}\left( 1 - \beta A^{[n]}_{ki}\frac{ \langle S_i^nI_k^n\rangle}{\langle S_i^n\rangle}\right).
\end{equation}

However, we run into the problem that we have no description for pairs of vertices. Thus, we derive expressions for their evolution from the RMEs for pairs of vertices which is given by,
\begin{equation}
\langle X_i^{n+1}X_j^{n+1} \rangle = \sum_{(X_{i}^n,X_{j}^n)\in \Omega^2}\langle X_i^{n+1}X_j^{n+1}|X_i^nX_j^n \rangle
\langle X_i^nX_j^n \rangle,
\end{equation}
For $\langle S_i^{n+1}I_j^{n+1} \rangle $, we obtain
\begin{align}\label{eq:SIR:TPB:SI}
\begin{split}
\langle S_i^{n+1}I_j^{n+1} \rangle =& \langle S_i^{n}I_j^{n}\rangle+
\langle S_i^{n+1}I_j^{n+1}|S_i^nS_j^n \rangle
\langle S_i^nS_j^n \rangle \\
&-  \langle I_i^{n+1}I_j^{n+1}|S_i^nI_j^n \rangle
\langle S_i^nI_j^n \rangle \\
&-  \langle S_i^{n+1}R_j^{n+1}|S_i^nI_j^n \rangle
\langle S_i^nI_j^n \rangle\\
&-  \langle I_i^{n+1}R_j^{n+1}|S_i^nI_j^n \rangle
\langle S_i^nI_j^n \rangle.
\end{split}
\end{align}
Note that the RME for $\langle S_i^{n}S_j^{n} \rangle$ is also required, which we find to be the following 
\begin{align}\label{eq:SIR:TPB:SS}
\begin{split}
\langle S_i^{n+1}S_j^{n+1}\rangle =& \langle S_i^nS_j^n\rangle -\langle S_i^{n+1}I_j^{n+1}|S_i^nS_j^n \rangle
\langle S_i^nS_j^n\rangle\\
&-\langle I_i^{n+1}I_j^{n+1}|S_i^nS_j^n \rangle
\langle S_i^nS_j^n\rangle \\
&-\langle I_i^{n+1}I_j^{n+1}|S_i^nS_j^n \rangle
\langle S_i^nS_j^n\rangle.
\end{split}
\end{align}

Since only the probabilities $\langle S_i^nI_j^n \rangle$ and $\langle S_i^nS_j^n \rangle$ are needed in order to describe the RMEs in Eq.~\eqref{eq:SIR:TPB:S:I}, we consider those two combinations of states. From \cite{frasca_discrete-time_2016}, we obtain the exact transition rates for pairs of vertices, though we find we can factorise the pair-wise transition rates similar to Eq.~\eqref{eq:SIR:S:I}. Here, we give the expression for $\langle S_i^{n+1}I_j^{n+1}|S_i^nS_j^n \rangle$ only while the rest of the pair-wise transition rates are given in Appendix~\ref{sec:app_A}:
{\footnotesize
	\begin{align}\label{eq:SIR:TPB:SS:SI}
		\langle S_i^{n+1}&I_j^{n+1}|S_i^nS_j^n \rangle = \nonumber\\
		& \frac{1}{\langle S_i^nS_j^n\rangle}\left[\beta\sum_{k_1\in V}A_{ik_1}^{[n]} \langle S_i^n S_j^nI_{k_1}^n\rangle\right.\nonumber
		- \beta^2\sum_{k_1, k_2\in V} A_{ik_1}^{[n]}A_{ik_2}^{[n]}\langle S_i^nS_j^n I_{k_1}^n I_{k_2}^n\rangle \nonumber\\
		& +\dots \nonumber\left.- (-\beta)^{N-2}\smashoperator{\sum_{k_1,\dots,k_{N-2}\in V}} A_{ik_1}^{[n]} \dots A_{ik_{N -2}}^{[n]}\langle S_i^n S_{j}^n \dots I_{k_{N-2}}^n\rangle\right]\nonumber\\
		&\times\left[1-\beta\sum_{k_1\in V}A_{ik_1}^{[n]} \langle S_i^n S_j^nI_{k_1}^n\rangle\right.
		+ \beta^2\sum_{k_1, k_2\in V} A_{ik_1}^{[n]}A_{ik_2}^{[n]}\langle S_i^nS_j^n I_{k_1}^n I_{k_2}^n\rangle \nonumber\\
		&-\dots \left.+ (-\beta)^{N-2}\smashoperator{\sum_{k_1,\dots,k_{N-2}\in V}} A_{ik_1}^{[n]}\dots A_{ik_{N -2}}^{[n]}\langle S_i^n S_{j}^n \dots I_{k_{N-2}}^n\rangle\right].
	\end{align}}

In the above equation, the term in the first pair of square brackets corresponds to the probability that vertex $i$ does not become infected and the term in the second pair of square brackets corresponds to the probability that vertex $j$ becomes infected. Upon applying our moment closure technique Eq.~\eqref{eq:SIR:TPB:SS:SI} may be written as
\begin{align}
\langle S_i^{n+1}&I_j^{n+1}|S_i^nS_j^n \rangle =  \nonumber\\
&\prod_{\substack{k\in V\\k\neq j}}
\left( 1-
\beta A^{[n]}_{ki}\frac{\langle S_i^nI_k^n \rangle}
{\langle S_i^n\rangle}\right)
\left [ 1 - 
\prod_{\substack{k\in V\\k\neq i}
}\left(1-
\beta A^{[n]}_{kj}\frac{\langle S_j^nI_k^n \rangle}
{\langle S_j^n\rangle}\right )\right].
\end{align}

By introducing the following functions, the RMEs for pairs as well as the individual vertices can be written more concisely. The probability that vertex $i$ does not become infected at time step $n+1$, given that $i$ is not infected at time step $n$ is denoted by
\begin{equation}
\Psi_i^n=\prod_{k\in V}\left( 1 - \beta A^{[n]}_{ki}\frac{ \langle S_i^nI_k^n\rangle}{\langle S_i^n\rangle}\right).
\end{equation}
Similarly, the probability that vertex $i$ does not become infected at time step $n+1$, given that $i$ is not infected at time step $n$ while excluding any interaction with $j$, is given by
\begin{equation}
\Phi_{ij}^n=\prod_{\substack{k\in V\\k\neq j}}\left( 1 - \beta A^{[n]}_{ki}\frac{\langle S_i^nI_k^n\rangle }{\langle S_i^n\rangle }\right).
\end{equation}
Then, the evolution of the state of every vertex in the network is determined by the following closed set of equations,
\begin{subequations}\label{eq:SIR:TPB:rme}
\begin{align}
\langle S_i^{n+1} \rangle =& \Psi_i^n\langle S_i^n \rangle\label{eq:SIR:TPB:S}\\
\langle I_i^{n+1}\rangle =& (1- \mu)\langle I_i^n\rangle + [1 - \Psi_i^n]\langle S_i^n\rangle\\
\langle S_i^{n+1}I_j^{n+1}\rangle =&(1-\mu) (1-\beta A_{ji})\Phi_{ij}^n\langle S_i^nI_j^n\rangle\nonumber \\
& + \Phi_{ij}^n(1-\Phi_{ji}^n)\langle S_i^nS_j^n\rangle\\
\langle S_i^{n+1}S_j^{n+1}\rangle =&\Phi_{ij}^n\Phi_{ji}^n \langle S_i^nS_j^n\rangle.
\end{align}
\end{subequations}

This approximation allows a large increase in accuracy compared to TIB model while only adding two equations to the final model. All past dynamic correlations are now tracked by the model and so the echo chamber effect is eliminated, but only with direct neighbours (vertices which share an edge). A major benefit of this particular TPB model over other existing iterations \cite{koher_contact-based_2019}\cite{gomez_discrete-time_2010} is that this model can be implemented as an element-wise sparse matrix multiplication rather than having to iterate through all edges for every time step making it extremely computationally efficient and fast on even large networks. It also benefits from a low conceptual cost by not deviating from a vertex-based perspective, like the contact-based models, which move to the perspective of edges and thus define the model in terms of the line-graphs and non-backtracking matrices\cite{koher_contact-based_2019}.\\

Similar to \cite{shrestha_message-passing_2015}, we can compare pair-based models to the TIB model using the two vertex example. In that illustrative configuration, we consider two vertices connected by an undirected static edge and give the two vertices some initial non-zero probability $\langle I_0^0\rangle = \langle I_1^0\rangle = z$ of being infected. We then run the TIB and TPB models for some given parameters $\beta$ and $\mu$ and compare it to the ground truth, which is the average of a number of Monte-Carlo realisations.

\begin{figure}[h]
	\centering
	\begin{tikzpicture}
	
	\draw (0,0) node[circle,minimum size=0.7cm,draw] (A) {\footnotesize $\langle I^0_0 \rangle=0.2$};
	\draw (3.5,0) node[circle,minimum size=0.7cm,draw] (B) {\footnotesize $\langle I^0_1 \rangle=0.2$};
	
	% Arrows
	\draw (A.east) -- (B.west);
	\end{tikzpicture}
\end{figure}

\begin{figure}[h]
	\centering
	\includegraphics[width=80mm]{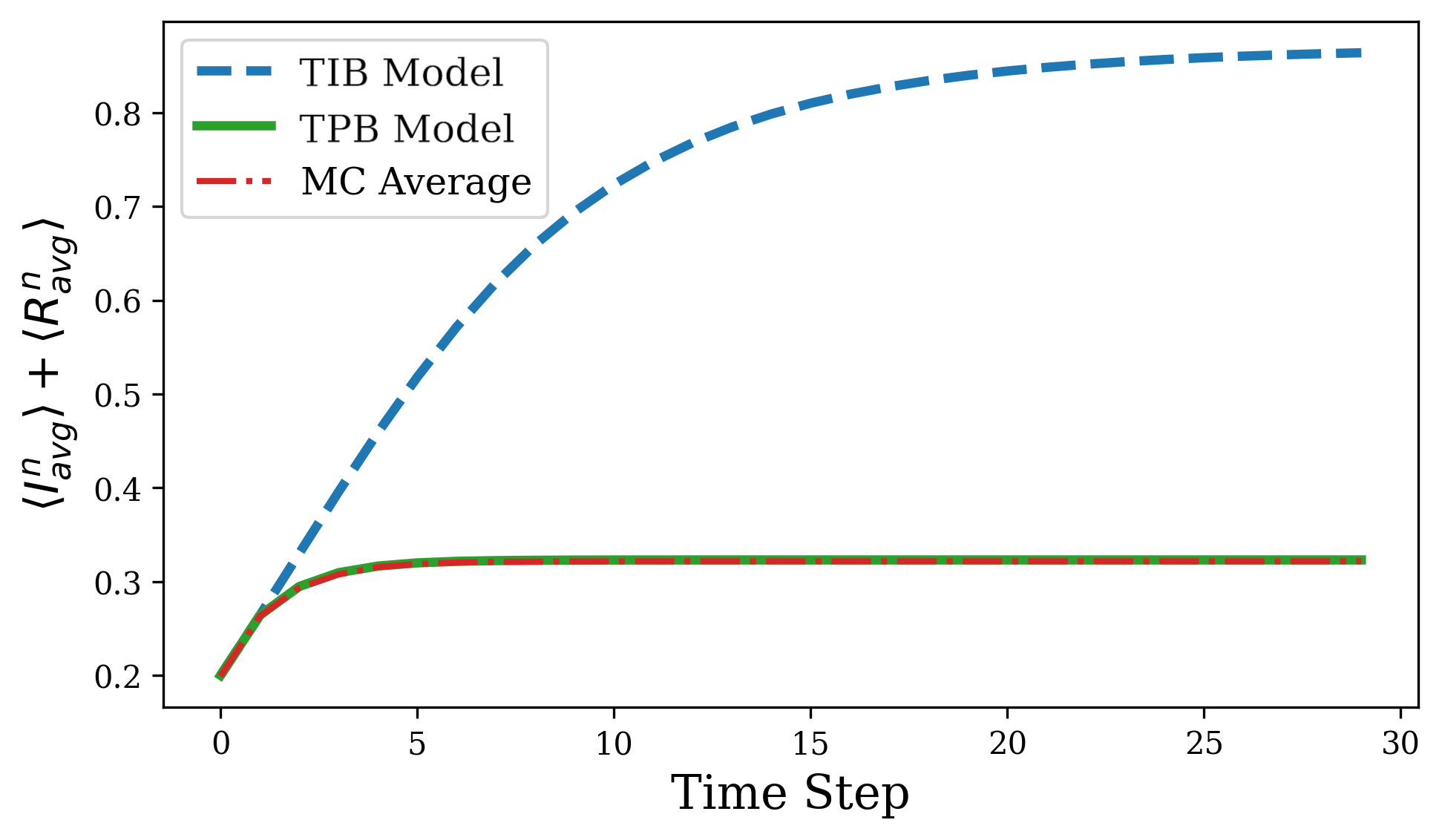}
	\caption{Running 25 time steps of the TIB and TPB SIR model as well as the average over 10$^5$ Monte-Carlo simulations for the two vertex example. Parameters: $\beta = 0.4$, $\mu = 0.2$ and $\langle I^0_0 \rangle= \langle I^0_1 \rangle=0.2$.}
	\label{fig:echo}
\end{figure} 

From Fig.~\ref{fig:echo}, it becomes apparent how the TIB model fails to capture the true SIR process on the network due to the previously discussed echo chamber induced by assuming statistical independence of vertices. It becomes clear that the TPB model accurately describes the underlying SIR process for this simple example as each vertex is able to recover the dynamic correlations of past interactions with direct neighbours.

\subsubsection{Equivalence Between The Contact-based and Pair-based Models}
In the contact-based model \cite{koher_contact-based_2019}, the central component is $\theta_{ij}^n$, which is the probability that node $j$ has not passed infection to node $i$ up to time step $n$. From $\theta_{ij}^n$, the quantity $\langle S^{n}_i\rangle$ may be computed as 
\begin{equation}\label{eq:SIR:CB:S}
\langle S_i^{n+1}\rangle = \langle S_i^0\rangle\prod_{j\in V}\theta_{ij}^{n+1}.
\end{equation}
This equation is the basis for the contact-based model and allows us to easily compare with the pair-based model as it describes the same quantity as our Eq.~\eqref{eq:SIR:TPB:S}. The authors also assume that the evolution of $\theta_{ij}^n$ satisfies the following relation
\begin{align}\label{eq:SIR:CB:theta}
\theta_{ij}^{n+1}= \theta_{ij}^n &- \beta A^{[n]}_{ji}\frac{\langle S_i^nI_j^n\rangle}{\langle S_i^n\rangle}\\
\theta_{ij}^{0}&=1.\nonumber
\end{align}

In the pair-based model, the evolution of the susceptible probability, given by Eq.~\eqref{eq:SIR:TPB:S}, can similarly be rewritten in terms of its initial conditions,
\begin{subequations}\label{eq:SIR:TPB:S:init_conds}
\begin{align}
\langle S_i^{n+1}\rangle &= \Psi_i^n\langle S_i^n\rangle\\
&= \langle S_i^0\rangle \prod^{n}_{m=0}\Psi_i^m\\
&= \langle S_i^0\rangle\prod_{j\in V}\prod^{n}_{m=0}
\left(1 - \beta A^{[m]}_{ji}\frac{\langle S_i^mI_j^m\rangle}{\langle S_i^m\rangle}\right).
\end{align}
\end{subequations}
From equating \eqref{eq:SIR:CB:S} and \eqref{eq:SIR:TPB:S:init_conds} it is clear that if the models are exactly equivalent then $\theta_{ij}$ is defined by
\begin{equation}\label{eq:SIR:TPB:theta}
\theta_{ij}^{n+1} = \prod^{n}_{m=0}\left(1 -\beta A^{[m]}_{ji}\frac{\langle S_i^mI_j^m\rangle}{\langle S_i^m\rangle}\right).
\end{equation}

However, this contradicts the assumption made by Eq.~\eqref{eq:SIR:CB:theta}. Thus the pair-based and contact-based models are only equivalent if the following linearisation is assumed:
\begin{equation}\label{eq:SIR:TPB:theta:approx}
\prod^{n}_{m=0}\left(1 - \beta A^{[m]}_{ji}\frac{\langle S_i^mI_j^m\rangle}{\langle S_i^m\rangle}\right)\approx 1 - \sum^n_{m = 0}\beta A^{[m]}_{ji}\frac{\langle S_i^mI_j^m\rangle}{\langle S_i^m\rangle},
\end{equation}
which then implies \eqref{eq:SIR:TPB:theta} can be written as
\begin{subequations}
\begin{align}
\theta_{ij}^{n+1} &= 1 - \sum^{n-1}_{m = 0}\beta A^{[m]}_{ji}\frac{\langle S_i^mI_j^m\rangle}{\langle S_i^m\rangle} - \beta A^{[n]}_{ji}\frac{\langle S_i^nI_j^n\rangle}{\langle S_i^n\rangle}\\
&= \theta_{ij}^n - \beta A^{[n]}_{ji}\frac{\langle S_i^nI_j^n\rangle}{\langle S_i^n\rangle}
\end{align}
\end{subequations}
This shows that the contact-based model is a linearised version of the pair-based model.

\section{Epidemic Threshold}\label{sec:epi_thresh}
One of the most important metrics used in epidemiology is the epidemic threshold, which determines the critical values of the model parameters at which a transition in qualitative behaviour occurs and the disease-free equilibrium (DFE), which we define as 

\begin{equation}\label{eq:DFE}
	\langle S_i^{n} \rangle = S_i^*,\quad
	\langle I_i^n\rangle = 0,\quad
	\langle R_i^n \rangle = R_i^*\quad \forall i
\end{equation}

where $\langle S_i^*\rangle + \langle R_i^*\rangle = 1$. From this definition it is clear that there exists a whole class of DFEs which must considered. At this critical point, the DFE becomes unstable and on average an epidemic occurs. Calculating the epidemic threshold is a bit more difficult with the SIR model compared with the SIS due to the fact that the flow of probability is in only one direction between compartments $S\rightarrow I \rightarrow R$, thus the class of DFE solutions are always asymptotically stable. Thus, we will look at classifying the initial stability of the SIR model as we perturb it from the state, 

\begin{equation}\label{eq:PDE}
	\langle S_i^{n} \rangle = 1,\quad
	\langle I_i^n\rangle = 0,\quad
	\langle R_i^n \rangle = 0\quad \forall i
\end{equation}

which we shall define as the pre-disease equilibrium (PDE). If it is unstable that means the disease has a chance to take hold and will spread through the network causing an epidemic before dying out. We now look at small perturbations from the PDE, if they vanish then the disease will die out and won't have a chance to propagate through the network. We shall define an epidemic in the SIR model as instability of the PDE under such perturbations. First, we look to linearise the difference equation for $\langle I_i^{n+1}\rangle$ near the PDE, this translates to linearising the non-linear function $\Psi_i^n$. Under the assumption $\langle I_i^n\rangle = \epsilon_i$ for every vertex $i$ such that $0<\epsilon_i\ll 1,$ we find 

\begin{equation}
		0\leq
		\frac{\langle S_i^nI_j^n\rangle}{\langle S_i^n\rangle}
		\leq\epsilon_j
\end{equation}

by the fact that for the joint probability 
$\langle S_i^nI_j^n\rangle\leq\min\{\langle S_i^n\rangle, \langle I_j^n\rangle\}$.
Thus for $\epsilon_i\ll 1$ we may assume 
$\frac{\langle S_i^nI_j^n\rangle}{\langle S_i^n\rangle}\approx\epsilon_j$. Upon substituting this into $\Psi_i^n$ we find that 

\begin{align}
\Psi_i^n
\approx &\prod_{k\in V}\left( 1 - \beta A^{[n]}_{ki}\epsilon_j\right)\nonumber\\
\approx &\sum_{k\in V}\beta A^{[n]}_{ki}\epsilon_j\label{eq:PSI:linear}
\end{align}

We can then use this to linearise $\langle I_j^{n+1}\rangle$ from \eqref{eq:SIR:TPB:rme}. While $\epsilon_i\ll 1$ holds, so does the approximation,
\begin{align}
\langle I_i^{n+1}\rangle & \approx  \langle I_i^{n}\rangle(1-\mu) + \sum_{k\in V} \beta A^{[n]}_{ik}\langle I_k^{n}\rangle.
\end{align}

This linearisation eliminates $\langle S_i^n\rangle$ from the equation. Interestingly, this is exactly the form of the SIS model in the TIB framework for which the epidemic threshold is easily found \cite{valdano_analytical_2015}. Therefore, we find that the SIS and SIR models share the same epidemic threshold condition. We introduce the matrix $M^{[n]}$, called the infection propagator, which is a linear map that describes the evolution of the SIR model close to the PDE:
\begin{equation}
M^{[n]}_{ij} = \beta A^{[n]}_{ij} + \delta_{ij}(1-\mu).
\end{equation}

Following Ref.~\cite{valdano_analytical_2015}, we find that the condition required for an epidemic to occur is given by
\begin{equation}
\rho\left( \prod_{m=0}^{n}{M}^{[m]} \right) > 1.
\end{equation}
For the values of $\beta$ and $\mu$ which the above is satisfied, implies that when a disease is introduced into the network the PDE is unstable for a period of time. What this means is that in the equivalent SIS model with the same parameters, the proportion of infected vertices never settles on the PDE.

\section{Results}\label{sec:results}
In this section, we compare the accuracy of the TIB model and the TPB model against the ground truth Monte-Carlo average, that is, direct stochastic simulations. In short, we show how the TPB model can offer a massive increase in accuracy and also discuss when it fails to accurately capture the true dynamics of the stochastic SIR process. Furthermore, we validate the analytical epidemic threshold.

\subsection{Tree Network}\label{sec:tree_network}
The assumption in the TPB model is conditional independence between vertices with a neighbour in common given the common neighbours state. This is equivalent to assuming the network contains no time-respecting non-backtracking cycles \cite{hashimoto1989zeta, bordenave2015non}. To illustrate this reasoning, we consider a static tree network, that is, tree networks contain no cycles of length 3 or greater, made up of 100 vertices. All vertices start from some initial non-zero probability $\langle I_i^0\rangle = z$ of being infected. We then run the TIB model and the TPB model for some given parameters $\beta$ and $\mu$ and compare it to the ground truth, which is the average of a number of Monte-Carlo realisations.

\begin{figure}[ht]
	\centering
	\includegraphics[width=0.9\linewidth]{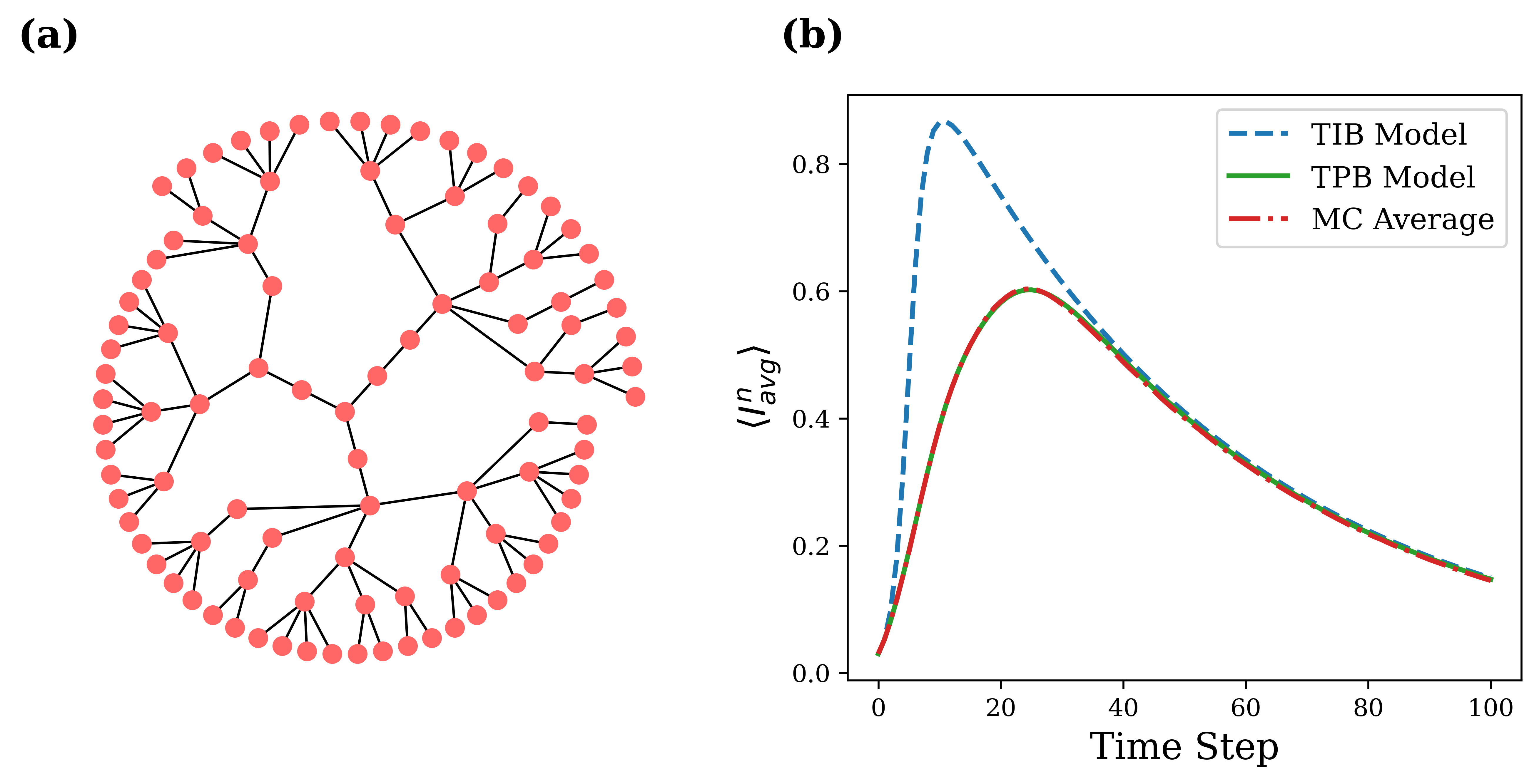}
	\caption{(a) A random tree network made up of 100 vertices. (b) Time series of the TIB (blue dashed) and TPB (green solid) SIR model as well as 10$^5$ Monte-Carlo (MC) simulations (red dotted) for the tree network shown in panel (a). Parameters: $\beta = 0.4$, $\mu = 0.02$ and $\langle I^0_0 \rangle= \langle I^0_1 \rangle=0.03$.}
	\label{fig:tree}
\end{figure}

Figure~\ref{fig:tree}(b) shows how the TIB model fails to capture the true SIR process on the network due to the previously discussed echo chamber induced by assuming statistical independence of vertices. It becomes clear that the TPB model accurately describes the underlying SIR process for this simple example as each vertex is able to recover the dynamic correlations of past interactions with direct neighbours. As we will see from the next section, temporal networks that are well approximated by tree networks are also well approximated by the TPB model.

\subsection{Empirical Networks}\label{sec:emp_networks}
In the following section, we consider 2 empirical temporal networks that all vary in both structure and temporal activity. For each of the empirical networks we wish to test our our findings that the TPB model offers an increase in accuracy over the TIB model. We observe a change in behaviour as the model parameters cross the epidemic threshold. We run the SIR TIB and TPB models for all our networks for different values of $\beta$ and $\mu$ and then compare them to the average of a sufficiently large number of MC simulations. This allows us to quantify how well the different models approximate the dynamics of the true SIR process. At each time-step, the average prevalence of states within the network are collected and denoted as $\langle S^n_{\text{avg}}\rangle$, $\langle I^n_{\text{avg}}\rangle$  and $\langle R^n_{\text{avg}}\rangle$ with the cumulative prevalence of infection being taken as $\langle I^n_{\text{avg}}\rangle + \langle R^n_{\text{avg}}\rangle$. Then, we validate our analytical finding for the epidemic threshold of the TPB SIR process. For this purpose, we fix a value for $\mu$ and then for increasing values of $\beta$, perform a number of MC simulations for long times in order to get a distribution of the final out break size, which is given by $ \lim_{n\rightarrow \infty}\langle I^n_{\text{avg}}\rangle + \langle R^n_{\text{avg}}\rangle$. In the long-term dynamics of the SIR process, $ \lim_{n\rightarrow \infty}\langle I^n_{\text{avg}}\rangle + \langle R^n_{\text{avg}}\rangle$,  will usually exceed the observation time of the network. Therefore, periodicity of the networks is assumed in a similar way to \cite{valdano_analytical_2015} when computing the final outbreak sizes.

\begin{table}[h]	
    \caption{List of empirical networks.}\label{tab:networks}
	\centering
	\footnotesize
\begin{tabular}{ |c|c|c|c|c|  }
	\hline
	\multicolumn{5}{|c|}{Network List} \\
	\hline
	Network& Vertices &Agg. Edges & Avg. Edges& Snapshots\\
	\hline
	Conference   & 405 & 9699 & 20.02 & 3509\\
	Cattle Trades&  111513  & 1041054   & 347.17 & 365\\
	\hline
\end{tabular}
\end{table}

\subsubsection{Irish Cattle Trade}\label{sec:cattle_network}

\begin{figure}[h!]
\includegraphics[width=0.8\linewidth]{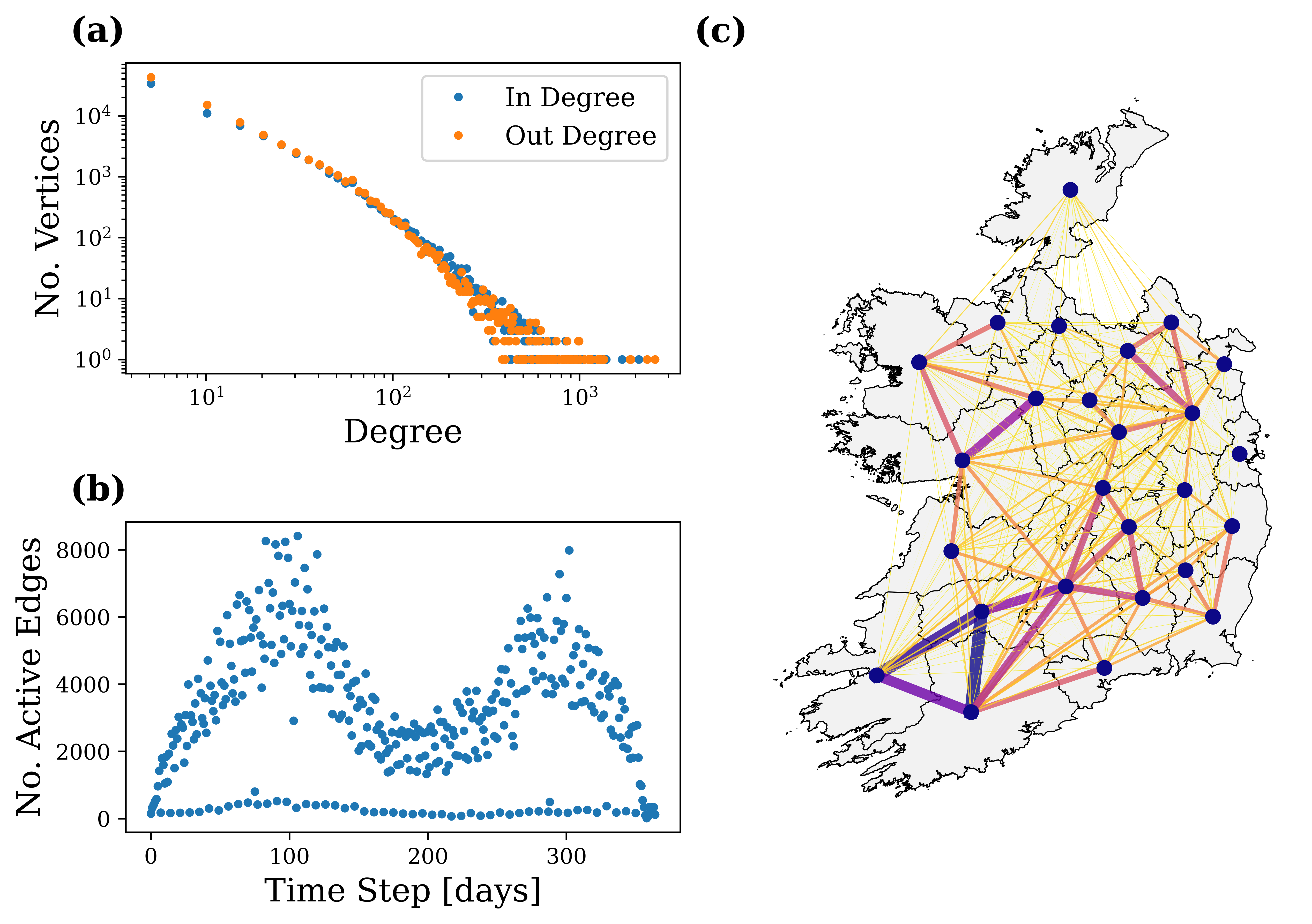}
	\caption{(a) In- and out-degree of the network aggregated over the entire year worth of data. (b) Time series of number of active edges per day in the network. (c) trades at the level of counties aggregated over the observation period. The number of trades is indicated by the edge width and colour.}
	\label{fig:cattle:data}
\end{figure}

The Cattle Trades network consists of all trades between herds within the Republic of Ireland during the year 2017 with a temporal resolution of one day \cite{TRATALOS2020105095}. Due to the nature of the trade data, interactions are directional. Thus, this data set is modelled by a directed network, where each vertex represents a herd and each edge represents a trade weighted by the number of animals traded. The aggregated degree distribution of the network as shown in Fig.~\ref{fig:cattle:data} (a) indicates a scale-free behaviour often seen in empirical networks. The network appears to be quite sparse as is evident from Fig.~\ref{fig:cattle:data} (b), with an average of only 347 edges per day while having an aggregated 1,041,054 edges over the entire year. The data also displays a strong bi-modal seasonal trend with there being two distinct peaks while there tends to be very little trades occurring on Sundays when the data points lie near zero. Although we ignore external drivers of the disease, this model still offers insight into how susceptible to epidemics the network is as trade is the main vector of non-local transmission. There are a number of infectious diseases that affect cattle, such as Foot and Mouth Disease and Bovine Tuberculosis, the latter of which is still a major problem in Ireland, thus effective models for the spread of infectious diseases among herds are incredibly useful tools. In the present study, we focus on the SIR dynamics, but the TPB model framework can easily be extended to other models,
%As described in the earlier sections we assume that a herd can belong to one of the compartments Susceptible, Infectious or Recovered. We assume the TPB SIR model and that no external factors influence the spread of disease (which is known to be the case however we will assume disease may only be spread through trade). 

\begin{figure}[h]
	\centering
	\includegraphics[width=0.8\linewidth]{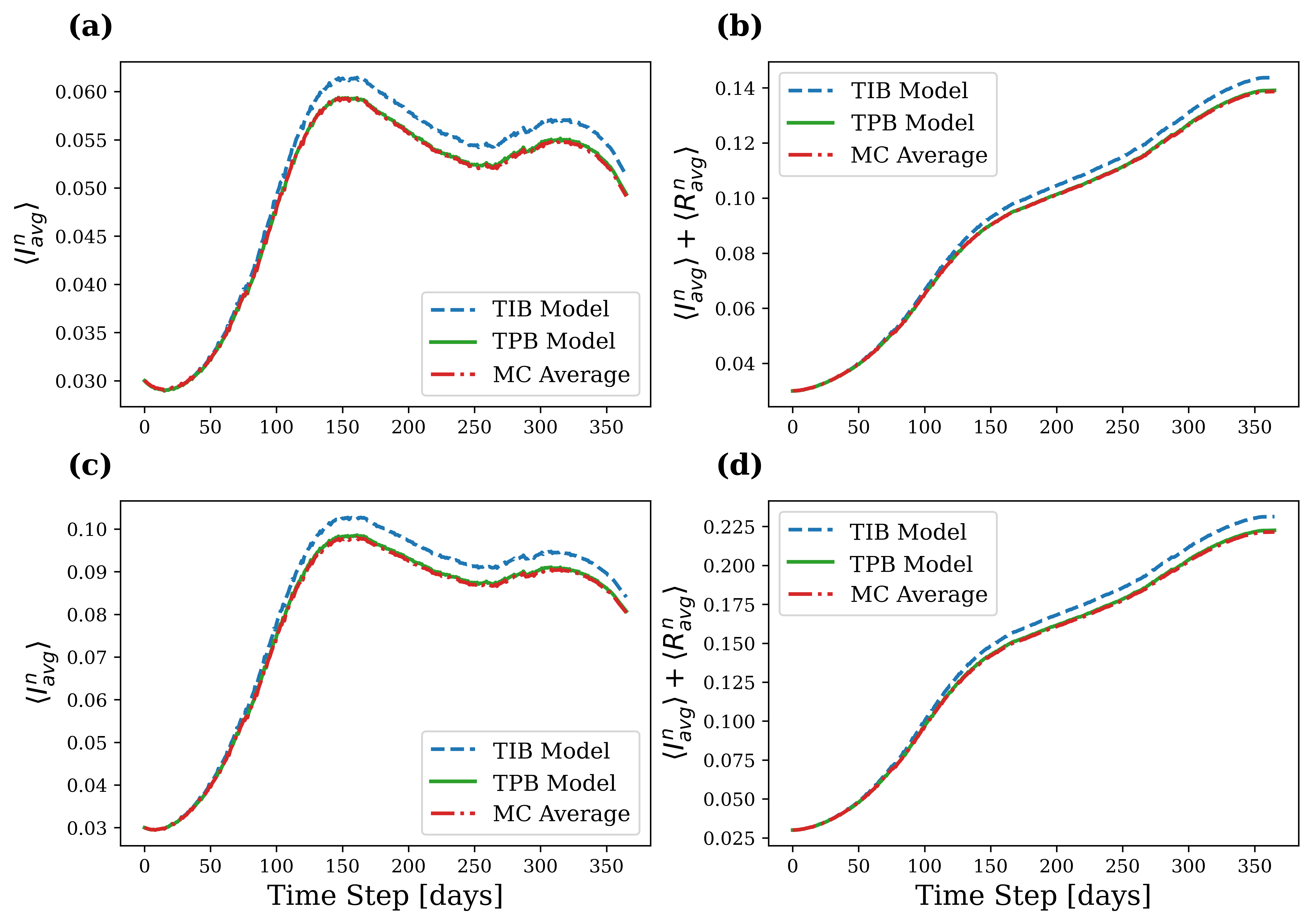}
	\caption{TIB (blue dashed) and TPB (green solid) SIR models on the Irish cattle trade network together with the average of $10^3$ Monte-Carlo simulations (red dotted). Panels (a),(c) show the time series for the prevalence of infection in the network and panels (b),(d) shows the cumulative prevalence of infection within the network. The probability of infection is set to 
	$\beta = 0.3$ in (a),(b) and 
	$\beta = 0.5$ in (c),(d). The initial chance of infection is 0.03. 
	Other parameters: $\mu=0.005$.}
	\label{fig:cattle:sir}
\end{figure}

\begin{figure}[h]
	\centering
	\includegraphics[width=0.95\linewidth]{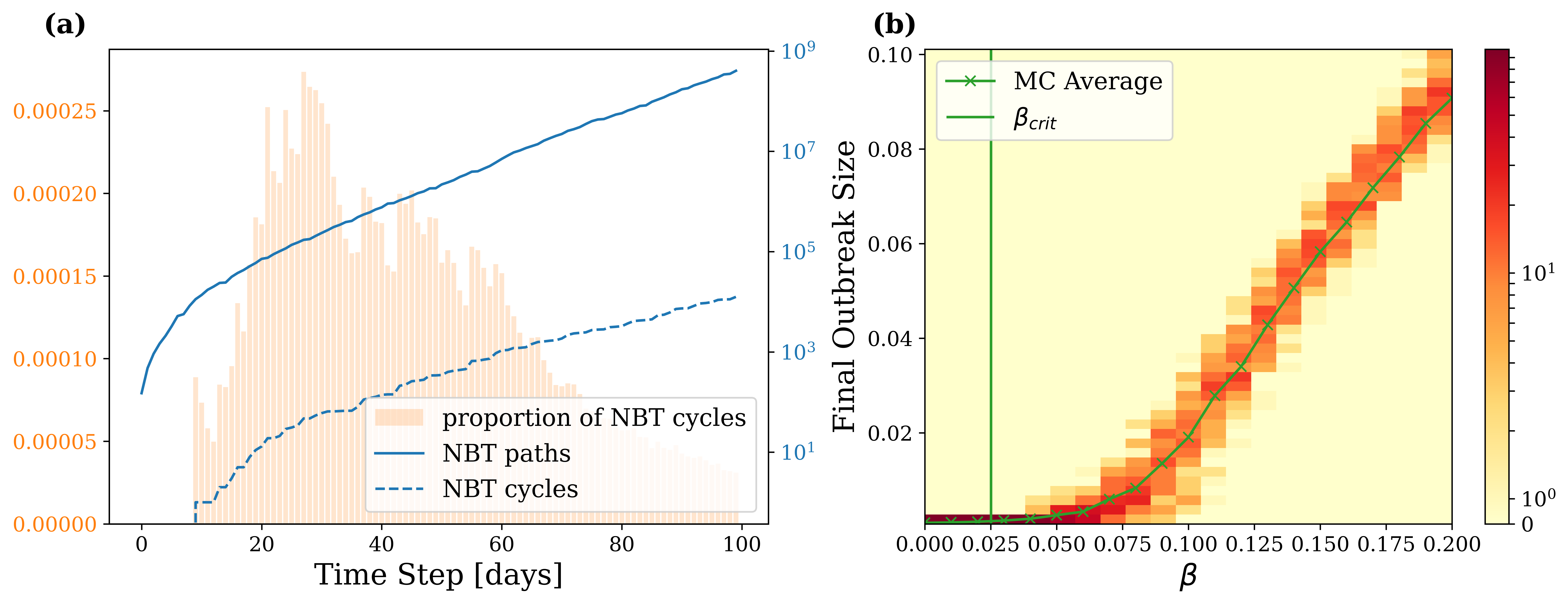}
	\caption{(a) Proportion (bars) and number of all non-backtracking (NBT) paths (blue solid) and cycles (blue dashed) that close at each respective time step. (b) Final outbreak sizes for varying values of $\beta$ with $\mu=0.005$. The vertical line shows the critical probability $\beta_{\text{crit}}$ obtained from the TPB model. The average of the Monte-Carlo (MC) is shown as green curve.}
	\label{fig:cattle:nbt_cycles}
\end{figure}

From Fig.~\ref{fig:cattle:sir} we can compare the performance of the TIB and TPB models on the cattle trades network. The figure shows a year worth of simulations of both models plotted against the average of $10^3$ MC realisations for the same choice of parameters. As is evident from the figure, the TPB model offers a significant improvement over the TIB model as there is a clear agreement with the MC average in the both the average and cumulative average disease prevalence for both choices of parameters. The reason for such a significant improvement can be explained by the fact that the TPB model is exact on networks with no non-backtracking cycles. However, because the cattle trade network is a production network, there exist very few non-backtracking cycles making the network structure highly tree-like in its supra-adjacency embedding. This can be explained by the fact that the existence of such cycles are inefficient and cost prohibitive in the trade process. As a result the network is well approximated by a tree network and contains very few non-backtracking cycles. Therefore, the SIR process is well approximated by the TPB model for such a network. As shown in Fig.~\ref{fig:cattle:nbt_cycles} the number of non-backtracking cycles as a fraction of the total number of paths in the network is very small, peaking at just over 0.025\%. Hence, the network is unlikely to suffer from the echo chamber effect in the TPB model. However, there are very many reciprocal (bi-directional) links in the network, meaning many farms trade in both directions. The reason for the difference in prediction between the TIB and TPB models is that the TPB model is able to account for all these reciprocal edges.

Next, we test our analytical findings for the epidemic threshold on the cattle trade network. Figure~\ref{fig:cattle:nbt_cycles}(b) depicts the average final outbreak sizes of a number of realisations against increasing values for $\beta$ while keeping $\mu$ fixed at 0.005. The critical $\beta$ which gives rise to an epidemic according to the analytically computed epidemic threshold for such a fixed $\mu$ in the TPB model is given as $\beta_{\text{crit}}\approx 0.025$. For values of $\beta$ that are greater than the computed epidemic threshold, there is an obvious but gradual change in dynamics as local outbreaks no longer die out, but now propagate throughout the network leading to larger final outbreak sizes as the value for $\beta$ gets larger, thus showing agreement with the analytical result for the epidemic threshold. Overall, we find that such trade networks are a good candidate for the TPB model as they avoid many non-backtracking cycles.

\subsubsection{Conference Contacts}
The second data set (cf. Tab.~\ref{tab:networks}) is the Conference network described in Ref.~\cite{genois_can_2018}. It includes the face-to-face interactions of 405 participants at the SFHH conference held in Nice, France 2009. Each snapshot of the network represents the aggregated contacts in windows of 20s. Since this data set describes face-to-face interactions, each contact is bi-directional and so an undirected network is the natural choice to model these interactions. 

\begin{figure}[h]
	\centering
	\includegraphics[width=0.8\linewidth]{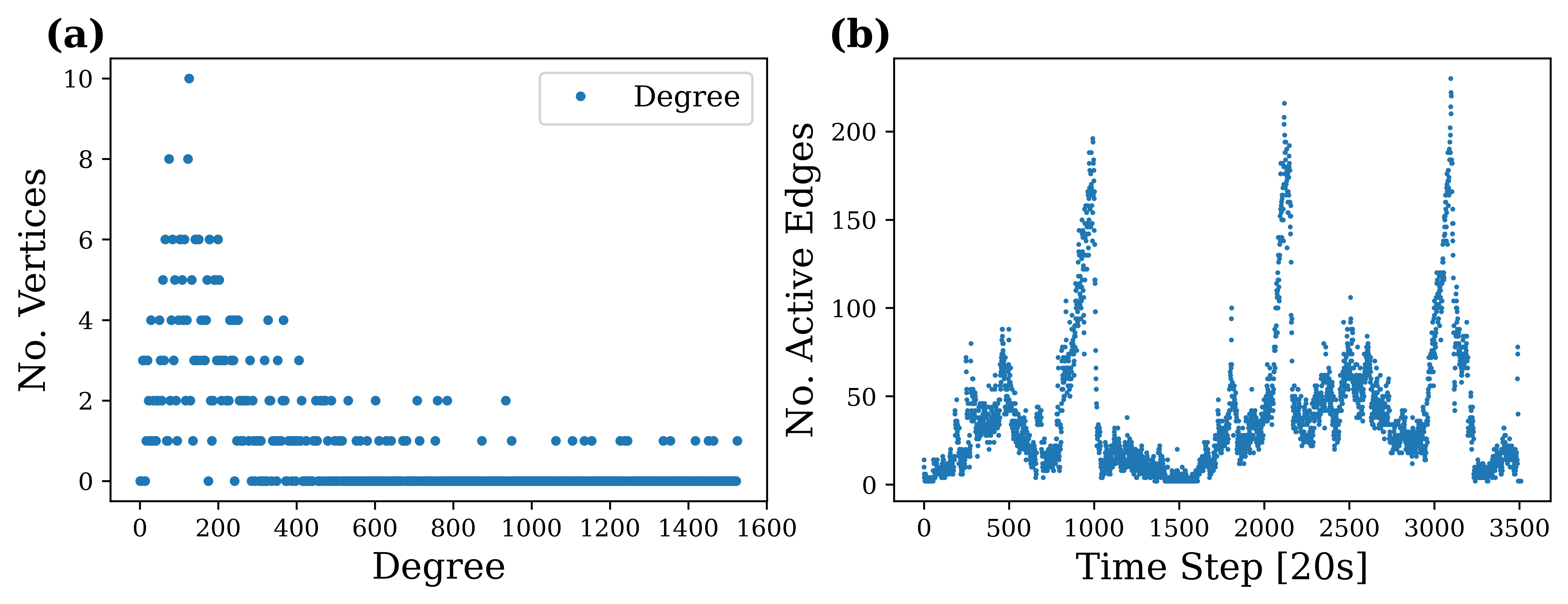}
	\caption{(a) Degree of the conference network aggregated over the entire observation period. (b) Time series of number of active edges per time step in the network.}
	\label{fig:conf:data}
\end{figure}

Because of the small number of nodes in the network, it is difficult to draw detailed conclusions from the degree distribution. As shown in Fig.~\ref{fig:conf:data}(a), there is a clear heavy tail with most vertices having a relatively small aggregated degree. In Fig.~\ref{fig:conf:data}(b) we see that the network activity in this case shows a number of peaks occurring then quickly dying out. These are explained by breaks between sessions at the conference during which the participants converse and interact. Because of the time scale and observation period of this particular temporal network, it is not feasible to model the spread of disease as infection and recovery is unlikely to occur within the observation period, which is approximately 20 hours. However, we can use our model to simulate the spread of viral information or "gossip" using the same dynamics as the SIR model. Infection is equivalent to receiving some information in such a way that it becomes interesting enough to for the individual to try and spread to those they contact in the future and recovery is equivalent to growing tired of the information and no longer inform others they meet.

\begin{figure}[h]
	\centering
	\includegraphics[width=0.8\linewidth]{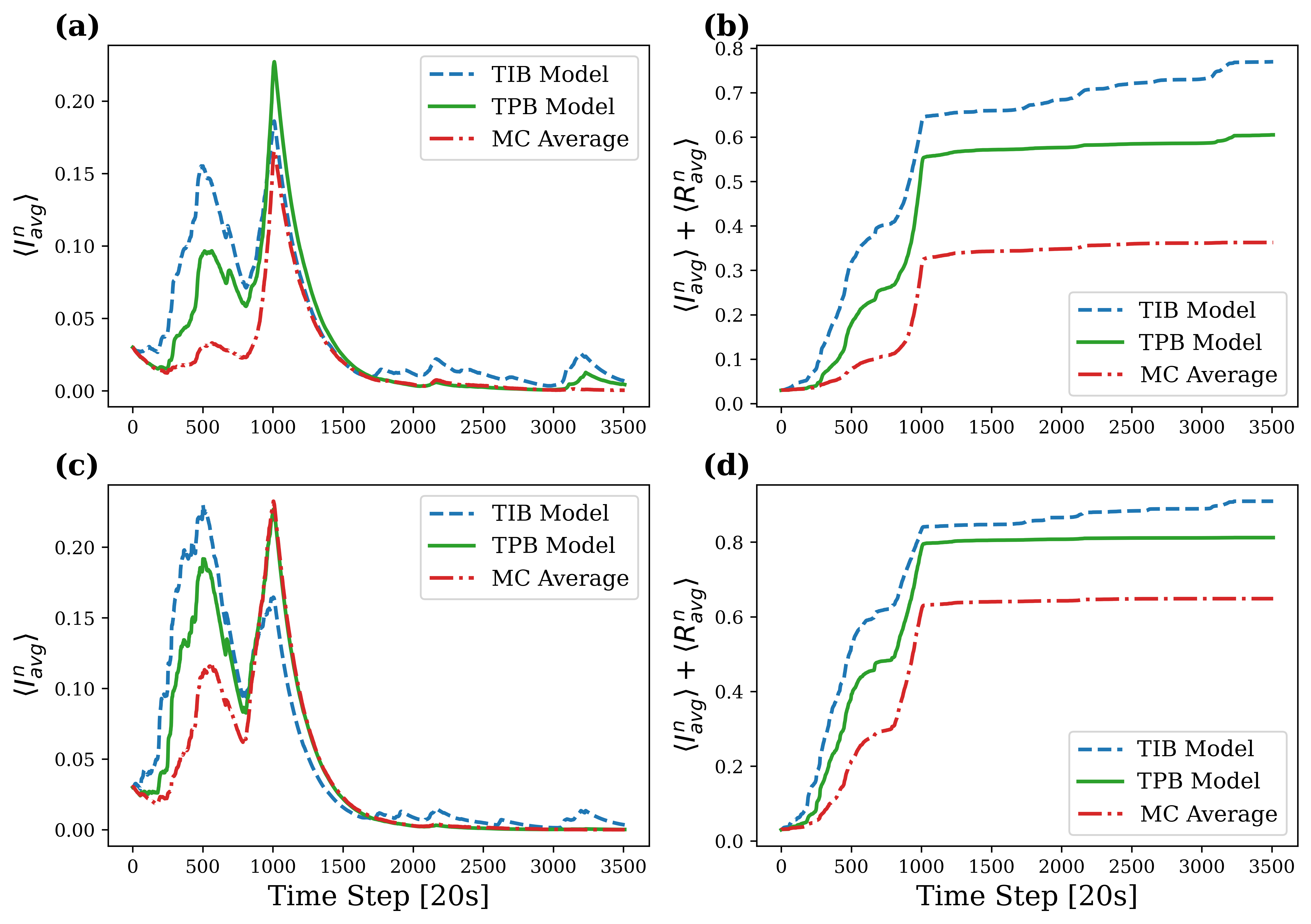}
	\caption{TIB (blue dashed) and TPB (green solid) SIR models on the conference network together with the average of $10^3$ Monte-Carlo (MC) simulations (red dotted). Panels (a),(c) show the time series for the prevalence of infection in the network. Panels (b),(d) depict the cumulative prevalence of infection within the network. 
	 The probability of infection is set to 
	$\beta = 0.3$ in (a),(b) and 
	$\beta = 0.5$ in (c),(d).
	The initial chance of infection is 0.03. Other parameters: $\mu=0.005$.}
	\label{model_test}
\end{figure}

Figure~\ref{model_test} shows the time series of the different models for two probabilities of infection: $\beta = 0.3$ in (a),(b) and 
$\beta = 0.5$ in (c),(d). Again, one can observe that in every case the TPB approximation offers an improvement over the TIB. However, compared to the Monte-Carlo simulations the TPB model does not offer perfect agreement in contrast to the cattle trade network. An interesting observation is that in panel (a) we see the average prevalence of the infection, according to the TPB model, has a higher peak than the TIB model and it may appear as though the TIB model performs better. However, when we look at the corresponding plot in panel (b), which shows the cumulative average prevalence of the disease, the TPB model is closer to the MC average at every time step. The reason the peak is bigger for the TPB model is that the TIB model has a sustained higher first peak, which uses up the pool of susceptible individuals leaving a smaller population available to catch the disease.\\

\begin{figure}[h]
	\centering
	\includegraphics[width=0.9\linewidth]{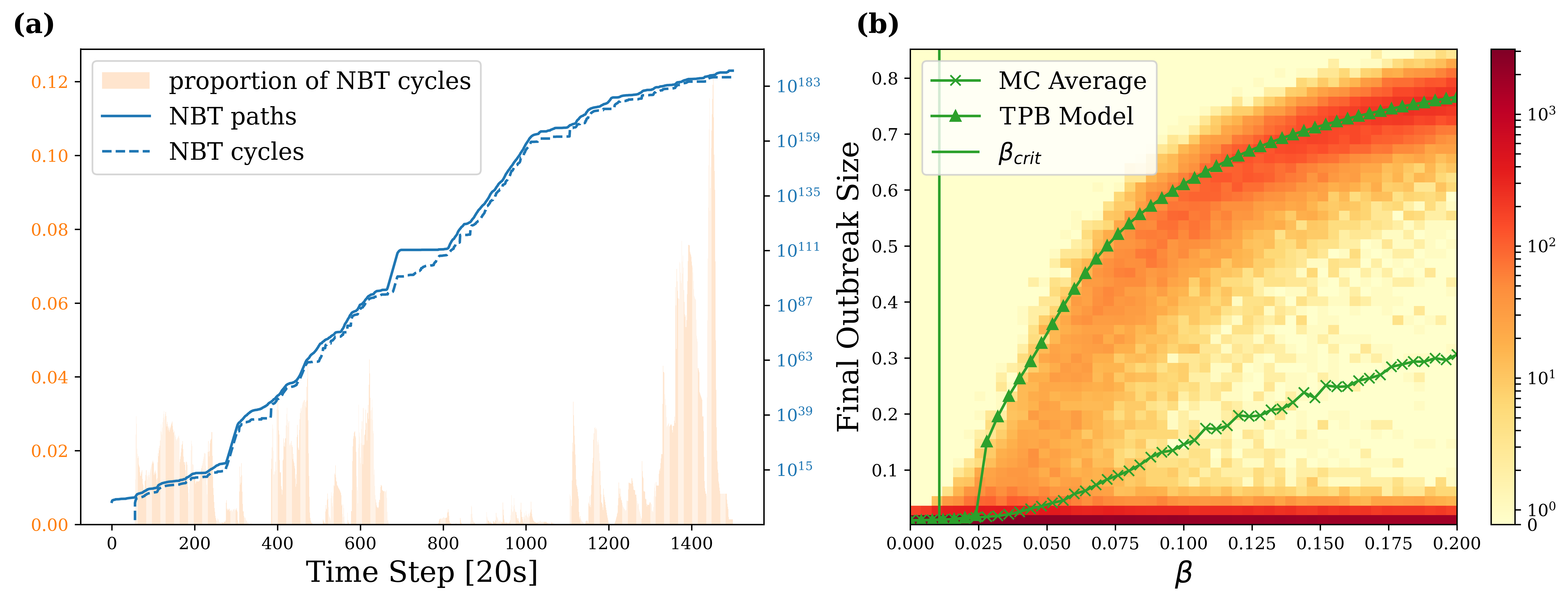}
	\caption{(a) Proportion (bars) and number of all non-backtracking (NBT) paths (blue solid) and cycles (blue dashed) that close at each respective time step. (b) Final outbreak sizes for varying values of $\beta$ with $\mu=0.005$.}
	\label{fig:conf:nbt_cycles}
\end{figure} 

The reason we do not see a good agreement with the MC average for this particular data set is due to the underlying topology of the network that is a physical social interaction network where individuals congregate in groups and most or all in the group interact with one another. This leads to large clusters that give rise to many non-backtracking cycles. The more time-respecting non-backtracking cycles that occur, the worse the TPB model will perform. It is for this reason that we see a relatively large deviation from the MC simulations for the TPB model. This can be explained by Figure~\ref{fig:conf:nbt_cycles}(a), which in contrast to the cattle network shows that the number of NBT-cycles is relatively dense at many points in time, reaching as high as 12\%.\\

In Fig.~\ref{fig:conf:nbt_cycles}(b) we see the distribution of final outbreak proportions against the critical $\beta$ computed for the epidemic threshold of the TPB model. Again, for values of $\beta$ that are greater than the computed epidemic threshold, there is an obvious but gradual change in dynamics as local outbreaks no longer die out, but now propagate throughout the network leading to larger final outbreak sizes as the value for $\beta$ gets larger, thus showing agreement with the analytical result for the epidemic threshold. However, the agreement with the final outbreak sizes only remains consistent with the MC average for values of $\beta$ below and slightly above the epidemic threshold.

\section{Conclusions}\label{sec:conclusions}

In this paper, we have presented work done on SIR pair-based models by systematically extending them to a temporal setting and investigating the effect of non-backtracking cycles on the accuracy of the model on arbitrary network structures. We have found that the existence of many such non-backtracking cycles leads to a deviation in the pair-based model from the true SIR process due to the echo chamber effect they induce. Thus, the pair-based model is best suited to network structures which do not contain many cycles, such as production networks. We also find that our analytical finding for the epidemic threshold holds up when compared to numerical simulations by, showing a qualitative change in the final outbreak proportion.

%%%%%%%%%%%%%%%%%%%%%%%%%%%%%%%%%%%%%%%%%%%%%%
%%                                          %%
%% Backmatter begins here                   %%
%%                                          %%
%%%%%%%%%%%%%%%%%%%%%%%%%%%%%%%%%%%%%%%%%%%%%%

\begin{backmatter}

\section*{Abbreviations}
\begin{description}
	\item [NBT] Non-Backtracking Tracking.
	\item [IB] Individual Based.
	\item [TIB] Temporal Individual-Based
	\item [PB] Pair Based.
	\item [TPB] Temporal Pair-Based.
	\item [MC] Monte-Carlo.
	\item [DFE] Disease-Free Equilibrium.
	\item [PDE] Pre-Disease Equilibrium.
\end{description}

\appendix

%\section{Exact Vertex Pair Transition Rates}

\section{Vertex Pair Transition Rates for the PB Model}\label{sec:app_A}
In this Appendix, we provide the transition rates used in Eqs.~\eqref{eq:SIR:TPB:SS:SI}:
{\footnotesize
    \begin{subequations}
	\begin{align}
	\langle S_i^{n+1}&I_j^{n+1}|S_i^nS_j^n \rangle = \nonumber \\
	&\prod_{\substack{k\in V\\k\neq j}}
	\left( 1-
	\beta A^{[n]}_{ki}\frac{\langle S_i^nI_k^n \rangle}
	{\langle S_i^n\rangle}\right)
	\left [ 1 - 
	\prod_{\substack{k\in V\\k\neq i}
	}\left(1-
	\beta A^{[n]}_{kj}\frac{\langle S_j^nI_k^n \rangle}
	{\langle S_j^n\rangle}\right )\right] 
	\end{align}
	\begin{align}
	\langle I_i^{n+1}&S_j^{n+1}|S_i^nS_j^n \rangle = \nonumber \\
	&\left [ 1 - 
	\prod_{\substack{k\in V\\k\neq j}}
	\left (1-
	\beta A^{[n]}_{ki}\frac{\langle S_i^nI_k^n \rangle}
	{\langle S_i^n\rangle}\right )\right] 
	\prod_{\substack{k\in V\\k\neq i}}\left (1-\beta A^{[n]}_{kj}\frac{\langle S_j^nI_k^n \rangle}{\langle S_j^n\rangle}\right)
	\end{align}
	\begin{align}
	\langle I_i^{n+1}&I_j^{n+1}|S_i^nI_j^n \rangle = \nonumber \\ 
	&(1-\mu)\left[1 - 
	(1-\beta A^{[n]}_{ji})\prod_{\substack{k\in V\\k\neq j}}
	\left (1-\beta A^{[n]}_{ki}
	\frac{\langle S_i^nI_k^n \rangle}
	{\langle S_i^n\rangle}\right )\right]
	\end{align}
	\begin{align}
	\langle S_i^{n+1}&R_j^{n+1}|S_i^nI_j^n \rangle = \nonumber \\
	&\mu\left [ (1 - \beta A^{[n]}_{ji}) 
	\prod_{\substack{k\in V\\k\neq j}}
	\left (1-\beta A^{[n]}_{ki}
	\frac{\langle S_i^nI_k^n \rangle}
	{\langle S_i^n\rangle}\right) \right]
	\end{align}
	\begin{align}
	\langle I_i^{n+1}&R_j^{n+1}|S_i^nI_j^n \rangle = \nonumber \\
	&\mu\left [ 1 - (1 - T_{ji}) 
	\prod_{\substack{k\in V\\k\neq j}}
	\left (1-\beta A^{[n]}_{ki}
	\frac{\langle S_i^nI_k^n \rangle}
	{\langle S_i^n\rangle}\right) \right].
	\end{align}
	\end{subequations}}
	
%%%%%%%%%%%%%%%%%%%%%%%%%%%%%%%%%%%
%%                               %%
%% Additional Files              %%
%%                               %%
%%%%%%%%%%%%%%%%%%%%%%%%%%%%%%%%%%%

%\section*{Additional Files}
%  \subsection*{Additional file 1 --- Sample additional file title}
%    Additional file descriptions text (including details of how to
%    view the file, if it is in a non-standard format or the file extension).  This might
%    refer to a multi-page table or a figure.

%  \subsection*{Additional file 2 --- Sample additional file title}
%    Additional file descriptions text.

\end{backmatter}
\end{document}